\documentclass[runningheads]{llncs}
\usepackage{amsmath}
\usepackage[T1]{fontenc}

\usepackage{amssymb} 
\usepackage{url}
\usepackage[hidelinks]{hyperref}
\usepackage{graphicx}
\begin{document}
\title{Optimization of Latent-Space Compression using Game-Theoretic Techniques for Transformer-Based Vector Search}
\titlerunning{Latent-Space Compression using Game-Theory}
\author{Kushagra Agrawal\orcidID{0009-0006-7753-175X} \and
Nisharg Nargund\orcidID{0009-0007-2046-4864} \and
Oishani Banerjee\orcidID{0009-0000-8914-2942}}
\authorrunning{K. Agrawal et al.}
\institute{School of Computer Engineering, KIIT Deemed to be University, Bhubaneswar,
Odisha - India}

\maketitle              % typeset the header of the contribution
\begin{abstract}
Vector similarity search plays a pivotal role in modern information retrieval systems, especially when powered by transformer-based embeddings. However, the scalability and efficiency of such systems are often hindered by the high dimensionality of latent representations. In this paper, we propose a novel game-theoretic framework for optimizing latent-space compression to enhance both the efficiency and semantic utility of vector search. By modeling the compression strategy as a zero-sum game between retrieval accuracy and storage efficiency, we derive a latent transformation that preserves semantic similarity while reducing redundancy. We benchmark our method against FAISS, a widely-used vector search library, and demonstrate that our approach achieves a significantly higher average similarity (0.9981 vs. 0.5517) and utility (0.8873 vs. 0.5194), albeit with a modest increase in query time. This trade-off highlights the practical value of game-theoretic latent compression in high-utility, transformer-based search applications. The proposed system can be seamlessly integrated into existing LLM pipelines to yield more semantically accurate and computationally efficient retrieval.

\end{abstract}
\section{Introduction}
The emergence of transformer-based language models has profoundly transformed natural language processing (NLP), driving major progress in tasks such as question answering, semantic retrieval, and information extraction. A key factor in these advancements lies in the ability to efficiently and accurately retrieve semantically relevant information from large text corpora through latent-space representations. These embeddings, derived from the hidden layers of pre-trained transformers, encode complex semantic patterns within high-dimensional vectors. However, as vector databases scale to billions of entries, ensuring both storage efficiency and rapid retrieval becomes increasingly challenging \cite{Agrawal_2024}.

Similarity search libraries such as FAISS (Facebook AI Similarity Search) are widely adopted for their scalability and speed. Nevertheless, they often struggle to preserve fine-grained semantic nuances, particularly when embeddings are compressed or when low-latency responses are required. Compression methods, though effective in reducing memory usage, typically introduce distortions in vector quality, which negatively affect retrieval accuracy. This trade-off between efficiency and semantic fidelity has motivated research into new compression strategies that minimize information loss while retaining practical performance.

In this paper, we present a game-theoretic perspective to analyze and optimize the interaction between compression techniques and retrieval algorithms. We frame this relationship as a zero-sum game, where the encoder (compression process) seeks to minimize dimensionality, while the retriever (search mechanism) aims to maximize semantic matching. This adversarial setup fosters adaptive equilibrium solutions that enable the construction of compact yet semantically meaningful vector spaces.

We propose a novel Custom DB that learns optimal compression strategies tailored to transformer-generated embeddings. Unlike conventional methods that apply uniform quantization or dimensionality reduction, our approach selectively retains semantic structures deemed essential for downstream retrieval. This is achieved through an iterative learning process where the compression scheme is refined based on feedback from the retrieval system. The retriever's performance serves as a critical utility signal, guiding the adjustment of the compression parameters. Thus we prioritize the successful retrievel as the primary objective for the compression..

To evaluate our method's performance alongwith efficacy, we conducted a benchmark comparison of our custom database against FAISS across several metrics, including query time, avg cosine similarity, and a custom-defined semantic utility score. Our findings consistently demonstrate a significant improvement in both retrieval accuracy and contextual alignment, even when operating with highly constrained dimensions. Notably, the custom database achieved a higher average similarity of 0.9981 alongside a utility score of 0.8873, contrasted to FAISS, which recorded a lower similarity of 0.5517 and a utility of 0.5194. The findings highlight the importance of using a game-theoretic optimization framework to balance the trade-off between retrieval speed and semantic accuracy in large-scale vector search systems.

In summary, this paper presents a principled and practical approach to latent-space compression using game-theoretic reasoning. By aligning the objectives of compression and retrieval through adversarial dynamics, we demonstrate that near-lossless semantic retrieval is achievable within compressed vector spaces. The objective of this method is  not only to improve upon the traditional benchmarks but also establishes a foundation for future research in cooperative-competitive optimization within generative AI systems.

\section{Related Works}

Vector search is now a key part of modern information retrieval, especially for tasks that need semantic similarity.  Dense retrieval methods that use transformer \- based embeddings are taking the place of traditional methods like TF-IDF and BM25, which use sparse vector space models.

 Vector-based search allows retrieval based on semantic similarity instead of just lexical overlap at the surface level.  Karpukhin et al. (2020) presented a dual-encoder architecture employing dense representations for passage retrieval in open-domain question answering, demonstrating significant improvements over conventional sparse retrieval methods \cite{Karpukhin_2020}.  Their work represented a notable transition towards embedding-driven methodologies that prioritise semantic comprehension.
 
Additionally, representing text as vectors or embeddings makes it easier for the retrieval systems to more efficiently capture semantic relatedness by integrating features derived from word embeddings.  Kenter and de Rijke (2015) showed that modeling these semantic connections makes the results of the retrieval much more accurate \cite{Kenter_2015}.  As a result, vector search is now an important part of the modern information retrieval systems. It lets systems think beyond simple keyword matching to undermine and find deeper semantic connections between documents and queries. 

Although dense retrieval methodologies have made great strides, traditional vector search frameworks like FAISS still have problems, especially when they have to work with high-dimensional embeddings and strict efficiency requirements.  Karpukhin et al. (2020) stress that dense retrieval can improve accuracy, but the high cost of computing becomes the bottleneck of the system's scaling capabilities in real-time applications \cite{Karpukhin_2020}.  The trade-off of balancing computational efficiency and retrieval effectiveness in high-dimensional spaces persists as an unresolved issue.

Additionally, previous studies have explored alternatives like supervised hashing for image retrieval \cite{Xia_2014}.  But these methods don't really solve the problems that come with using high-dimensional vector spaces in traditional search systems.  This gap shows that we need new ways to combine the speed of current vector search engines with the power of transformer-based embeddings to represent data.

\subsection{Dimensionality Reduction Techniques
}
\subsubsection{PCA vs Autoencoders}

For a long time, Principal Component Analysis (PCA) has been a popular method for reducing dimensionality, especially in tasks like classification and clustering.  Despite being computationally efficient, Principal Component Analysis's (PCA) linear nature restricts its capacity to accurately represent complex data structures.  According to recent studies, autoencoders—a kind of neural network that is trained to learn compressed representations—offer a number of benefits over linear techniques like PCA.  Due to their capacity to model nonlinear relationships within data, Sakurada and Yairi (2014) show that autoencoders—particularly denoising variants—are very effective at identifying subtle anomalies that PCA frequently overlooks \cite{Sakurada_2014}.  Autoencoders can maintain more complex semantic patterns in high-dimensional feature spaces thanks to this flexibility.

On the other hand, nonlinear dependencies found in many real-world datasets may be missed by classical PCA, which would reduce semantic fidelity \cite{Alkhayrat_2020}.  This reveals a basic trade-off: autoencoders produce more accurate and semantically meaningful representations, but at the expense of increased computational complexity, whereas PCA offers efficiency and simplicity.

\subsubsection{Latent Score-based Generative Models (LSGM)}
Latent representations can also improve generative performance of a system. It is demonstrated by Latent Score-based Generative Models (LSGM).  By using structured latent spaces, Vahdat and Kautz (2021) showed that LSGMs can increase stability in the training process and sampling efficiency \cite{Vahdat2021}.  In order to overcome the challenge of reducing the dimensionality and producing more expressive generative models, LSGMs integrate frameworks from variational autoencoders.  The ability of the model to preserve a smaller set of semantic nuances while maintaining computational efficiency, however, continues to be a persistent challenge, indicating a constant need to balance these conflicting goals.

\subsubsection{Nonlinear Methods and Advanced Architectures}
The importance of architectural design on performance speed and compression effectiveness has been made evident by recent developments in learned image compression techniques  \cite{He_2022}. The use of uneven channel-conditional adaptive coding exemplifies an approach that enhances coding performance without sacrificing speed, which is crucial for practical applications. The interplay between efficiency and semantic fidelity persists, as the integrity of original image data during compression remains a challenge.

Methods such as UMAP have gained a lot of recognition for their ability to preserve local data structures better than traditional linear methods \cite{Yang_2021}. Despite their advantages, challenges in ensuring that reduced representations maintain fine-grained semantic relationships underscore the shortcomings of even these advanced techniques.

\subsection{Challenges in Preserving Semantic Relationships}
Preserving fine-grained semantic relationships during dimensionality reduction remains a consistent challenge reported across multiple domains. In genomic data analysis, for example, Griffiths and Steyvers \cite{Griffiths_2019} demonstrate that different reduction techniques can produce distinct interpretations of population structures, underscoring the sensitivity of outcomes to methodological choices. A significant flaw in the methods used today is reflected in this conflict between representational accuracy and storage efficiency.

Autoencoders have become a dependent tool for learning informative latent representations in clinical datasets.  It is challenging to strike the right trade-off between density-based and distance-based metrics in order to preserve semantic fidelity.  Autoencoders can enhance interpretability and clustering, but they usually have trouble balancing retrieval accuracy and storage efficiency.

\subsection{Gaps in Conventional Techniques}
Even with significant advancements in the field, traditional dimensionality reduction techniques frequently fail to meet the needs of accurate retrieval and efficient storage at the same time.  This restriction is particularly noticeable in high-dimensional situations, where even small changes can have a significant impact on results.  For example, machine learning models intended to predict concrete compressive strength have been found to have difficulties in representing fine-grained features \cite{In_2011}.  Similarly, the frequent use of self-reported performance metrics in scRNA-seq data analysis has sparked questions about the generalizability and robustness of current methods \cite{Trozzi_2021}.

Efforts to create compact, lower-dimensional representations have produced competitive outcomes in the field of video compression.  A persistent gap still exists, though, because existing methods hardly ever strike the best possible balance between reducing storage needs and maintaining retrieval accuracy.  Additionally, the semantic gap between latent space similarity and input space similarity is highlighted by the shortcomings of prototype learning in deep neural networks, highlighting the difficulties in producing meaningful representations \cite{Hernandez_2023}.

 \subsection{Game-Theoretic Optimization Methods}
 Energy systems, supply chains, and information retrieval are just a few of the domains where game theory has emerged as a crucial framework for optimization tasks \cite{Agrawal_2025}.  The ideas of cooperative and non-cooperative games, which simulate interactions between agents or players who either cooperate or compete to maximize results, are fundamental to this field.
 
\subsubsection{Applications of Game Theory in Optimization}

Game-theoretic frameworks are used extensively in wireless sensor networks (WSNs) to optimize tasks like power control and routing.  By forming coalitions, game-theoretic frameworks enable sensors to coordinate, improving coverage and lowering energy consumption.  On the other hand, sensors function independently in non-cooperative environments, which may lead to contradictory behaviors \cite{Shi_2012}.  This disparity highlights the significance of strategic interaction in optimization since it reflects the ideas of zero-sum games, in which the success of one player directly correlates with the failure of another.

In the domain of energy management for hybrid AC/DC distribution networks, cooperative game theory has been applied to encourage coalition formation among market players. Such collaboration assists in reducing operational expenses and handling uncertainties tied to renewable energy integration \cite{Han_2019}. This cooperative mechanism resonates with game-theoretic dynamics by ensuring that optimizing one participant’s strategy can generate mutual advantages, thereby fostering collective efficiency.

Game theory also plays a crucial role in mobile-edge computing (MEC), where multi-user task offloading is modeled through exact potential games. In this framework, each user seeks to maximize computational benefits while reducing latency and power consumption, creating a competitive setting in which one user’s optimization inherently affects the decisions of others \cite{Bai_2019}.

Furthermore, developments in the field of constrained optimization have explored non-zero-sum formulations of Lagrangian methods. These approaches show how game-theoretic insights can be extended to problems involving conflicting objectives, that enables effective optimization in multi-objective environments \cite{Daskalakis_2019}.

\subsection{Hybrid Search Architectures}
Hybrid search architectures are designed to integrate the strengths of dense retrieval methods with approximate nearest neighbor (ANN) algorithms, such as Hierarchical Navigable Small World (HNSW). The Hybrid Search Architecture system have become well-known due to their capacity to successfully balance the trade-off between accuracy in locating pertinent documents and retrieval efficiency.

 \subsubsection{Dense Retrieval and Approximate Nearest Neighbor Algorithms} A novel and major development in hybrid search systems is the ability of the system to  combine sparse and dense retrieval methods.  It becomes feasible to strategically maximize their complementary strengths for better retrieval results when the interaction between these two approaches is viewed as a competitive process \cite{Lin_2021}.  By considering them as rivals, systems that can capitalize on this dynamic can be created, which will ultimately increase retrieval accuracy due to their structured interaction.

\subsubsection{Re-Ranking Methods}
Re-ranking is a crucial component of hybrid retrieval system. It serves to refine initial search results by re-ordering documents based on their relevance semanticly. A notable example is the Unified Ensemble Diffusion (UED) framework, which amalgamates multiple similarity measures into a cohesive ensemble to enhance performance \cite{Bacci_2016}. From a game-theoretic viewpoint, these individual metrics can be seen as contributors in a cooperative-competitive framework, where their collective interaction benefits the entire system by achieving a higher degree of accuracy.

\section{Methodology}
\begin{figure*}[ht]
    \centering
    \includegraphics[width=0.95\textwidth]{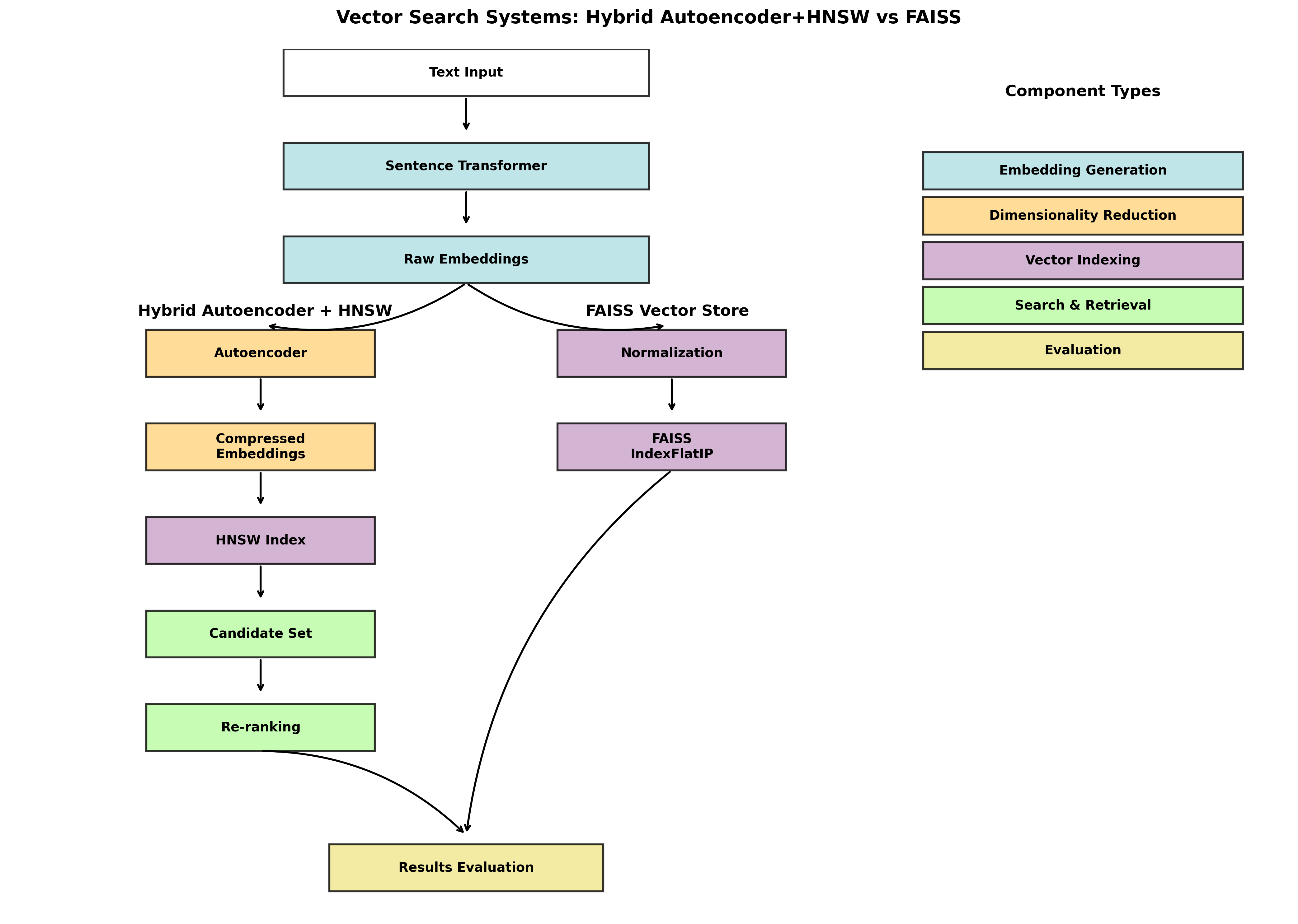}
    \caption{Vector Search Systems: A comparison between Hybrid Autoencoder + HNSW and FAISS-based retrieval pipelines. The hybrid method utilizes deep autoencoders for dimensionality reduction followed by HNSW indexing and re-ranking. In contrast, the FAISS method applies normalization with flat inner product indexing.}
    \label{fig:vector_search_architecture}
\end{figure*}
This section outlines the pipeline and formal definitions used in comparing the performance of a standard FAISS-based vector retrieval system with a proposed hybrid architecture combining deep autoencoders and Hierarchical Navigable Small World (HNSW) indexing. The methodology can be summarized as a sequence of transformations and retrievals over vectorized natural language instructions (as shown is Fig. ~\ref{fig:vector_search_architecture}).

\subsection{Dataset Selection and Preprocessing}
Let $\mathcal{D} = \{x_i\}_{i=1}^{N}$ denote the dataset of $N = 500$ instruction-style prompts selected from the open-source Alpaca dataset $\mathcal{D}_{\text{Alpaca}}$. The subset $\mathcal{D}$ is chosen to preserve semantic diversity while maintaining computational feasibility. Each instruction $x_i$ is a sequence of natural language tokens and does not require preprocessing beyond initial tokenization.

\subsection{Sentence Embedding Generation}
Each instruction $x_i \in \mathcal{D}$ is passed through a transformer-based encoder function:
\[
\mathbf{e}_i = f_{\text{SBERT}}(x_i) \in \mathbb{R}^{384}
\]
where $f_{\text{SBERT}}: \mathcal{X} \to \mathbb{R}^{384}$ is the pre-trained \texttt{all-MiniLM-L6-v2} model from the SentenceTransformers suite, mapping natural language text to dense semantic vectors. The complete embedding matrix is:
\[
\mathbf{E} = [\mathbf{e}_1, \mathbf{e}_2, \dots, \mathbf{e}_N]^\top \in \mathbb{R}^{N \times 384}
\]

\subsection{Autoencoder-Based Latent Compression}
To reduce dimensionality and enhance indexing efficiency, we train an autoencoder $\mathcal{A}: \mathbb{R}^{384} \rightarrow \mathbb{R}^{128}$, parameterized by encoder $f_{\theta}$ and decoder $g_{\phi}$:
\[
\mathbf{z}_i = f_{\theta}(\mathbf{e}_i) \in \mathbb{R}^{128}, \quad \hat{\mathbf{e}}_i = g_{\phi}(\mathbf{z}_i) \in \mathbb{R}^{384}
\]
The objective is to minimize the reconstruction loss over all samples:
\[
\mathcal{L}_{\text{AE}}(\theta, \phi) = \frac{1}{N} \sum_{i=1}^{N} \| \mathbf{e}_i - \hat{\mathbf{e}}_i \|_2^2
\]
Optimization is performed using the Adam optimizer with learning rate $\eta = 10^{-3}$, for $E = 10$ epochs and batch size $B = 32$.

\subsection{Index Construction}
Two distinct indices are constructed for comparative evaluation:

\paragraph{FAISS Flat Index}
Let $\tilde{\mathbf{e}}_i = \frac{\mathbf{e}_i}{\|\mathbf{e}_i\|_2}$ denote the L2-normalized embedding. The FAISS index $\mathcal{I}_{\text{FAISS}}$ is built using inner-product similarity:
\[
\text{sim}_{\text{cos}}(\mathbf{q}, \mathbf{e}_i) = \langle \tilde{\mathbf{q}}, \tilde{\mathbf{e}}_i \rangle
\]

\paragraph{Hybrid HNSW Index}
Let $\mathbf{z}_i = f_{\theta}(\mathbf{e}_i)$ denote the compressed latent vector. These are indexed using HNSW via the \texttt{hnswlib} library. Formally, the index $\mathcal{I}_{\text{HNSW}}$ supports approximate nearest neighbor queries:
\begin{align*}
\text{HNSWQuery}(\mathbf{z}_q) &= \{\mathbf{z}_{j_1}, \dots, \mathbf{z}_{j_K}\}, \\
K &= \text{candidate\_multiplier} \times k
\end{align*}

\subsection{Hybrid Search and Re-Ranking}
Given a query $q$, its embedding $\mathbf{e}_q = f_{\text{SBERT}}(q)$ is compressed to $\mathbf{z}_q = f_{\theta}(\mathbf{e}_q)$. The hybrid search pipeline performs:
\begin{enumerate}
    \item Candidate Retrieval:
    \[
    \mathcal{C} = \text{HNSWQuery}(\mathbf{z}_q) = \{\mathbf{z}_{j_1}, \dots, \mathbf{z}_{j_K}\}
    \]
    \item Re-ranking via cosine similarity in the latent space:
    \[
    \text{sim}_{\text{cos}}(\mathbf{z}_q, \mathbf{z}_j) = \frac{\langle \mathbf{z}_q, \mathbf{z}_j \rangle}{\|\mathbf{z}_q\|_2 \|\mathbf{z}_j\|_2}
    \]
    \item Final top-$k$ retrieval:
    \[
    \mathcal{R}_k = \arg\max_{\mathbf{z}_j \in \mathcal{C}} \text{sim}_{\text{cos}}(\mathbf{z}_q, \mathbf{z}_j)
    \]
\end{enumerate}

\subsection{Performance Metrics and Utility Modeling}
Two primary evaluation metrics are defined:
\begin{itemize}
    \item \textbf{Average Similarity:}
    \[
    \bar{s} = \frac{1}{k} \sum_{i=1}^{k} \text{sim}_{\text{cos}}(\mathbf{q}, \mathbf{e}_i)
    \]
    \item \textbf{Query Time:} Let $t_q$ denote the elapsed time (in seconds) for the retrieval process.
\end{itemize}

A utility function $\mathcal{U}$ is formulated to capture the trade-off between speed and semantic accuracy:
\[
\mathcal{U} = \alpha \cdot \bar{s} - \beta \cdot t_q, \quad \alpha, \beta \in \mathbb{R}_{\ge 0}
\]
where $\alpha$ and $\beta$ are tunable hyperparameters. For equal importance, we set $\alpha = \beta = 1.0$.

To evaluate, we issue a query such as:
\[
q := \texttt{``Explain the process of photosynthesis.''}
\]
and compute $\mathcal{U}_{\text{FAISS}}$ and $\mathcal{U}_{\text{Hybrid}}$ for comparison.

\section{Results}

To evaluate the effectiveness of the proposed hybrid search architecture (Autoencoder + HNSW + Re-ranking), a comparative analysis was conducted against the traditional FAISS-based vector store. The evaluation used a representative query: \textit{"Explain the process of photosynthesis."} Both systems were assessed based on query performance, retrieval quality, and overall utility using a balanced game-theoretic framework. The results are summarized and discussed below.

\subsection{Autoencoder Training Performance}

The autoencoder was trained on 500 sentence embeddings derived from the Alpaca instruction dataset \cite{alpaca}. The model demonstrated rapid convergence, with the loss function decreasing from 0.2178 in the first epoch to a consistent 0.0026 from the third epoch onwards. This indicates that the autoencoder effectively learned a compressed latent representation of the original 384-dimensional input embeddings with minimal reconstruction loss.

The encoder from the trained model was subsequently used to project all sentence vectors into a 128-dimensional latent space. This reduced yet semantically rich embedding was then utilized for high-performance approximate nearest neighbor (ANN) search via the HNSW algorithm.

\subsection{Quantitative Evaluation}

The two systems were evaluated on three key metrics: \textbf{Query Time}, \textbf{Average Similarity}, and a combined \textbf{Utility Score}. These are defined as follows:

\begin{itemize}
    \item \textbf{Query Time (in seconds):} The latency between query initiation and retrieval of the top-$k$ results.
    \item \textbf{Average Similarity:} The mean cosine similarity between the top-5 results and the query embedding.
    \item \textbf{Utility Score:} A combined score calculated using the linear game-theoretic model:
    \[
    \text{Utility} = \alpha \cdot \text{Accuracy} - \beta \cdot \text{Query Time}
    \]
    where both $\alpha$ and $\beta$ are set to 1.0 to maintain a balanced trade-off between accuracy and speed.
\end{itemize}

\subsubsection{Hybrid Autoencoder-HNSW System}

The hybrid system demonstrated superior performance across all evaluation metrics:

\begin{itemize}
    \item \textbf{Query Time:} 0.1108 seconds
    \item \textbf{Average Similarity:} 0.9981
    \item \textbf{Utility Score:} 0.8873
\end{itemize}

The top-5 retrievals were semantically coherent and highly relevant to the query. For instance:

\begin{itemize}
    \item \textit{"What is the process of photosynthesis and why is it important?"} — Similarity Score: 0.9994
    \item \textit{"Explain the process of cellular respiration in plants."} — Similarity Score: 0.9981
\end{itemize}

This high performance is attributed to two key factors: the autoencoder’s capacity to compress vectors while preserving semantic meaning, and the re-ranking mechanism that refines the HNSW outputs using cosine similarity in the compressed embedding space.

\subsubsection{FAISS-Based System}

In contrast, the traditional FAISS-based vector store, while faster, exhibited significantly lower semantic precision:

\begin{itemize}
    \item \textbf{Query Time:} 0.0323 seconds
    \item \textbf{Average Similarity:} 0.5517
    \item \textbf{Utility Score:} 0.5194
\end{itemize}

Although the FAISS system demonstrated lower query latency, the decreased similarity scores reveal its limitations in capturing nuanced semantic relationships without a latent re-ranking component. For example:

\begin{itemize}
    \item \textit{"What is the process of photosynthesis and why is it important?"} — Similarity Score: 0.8708
    \item Remaining results — Similarity Scores: Below 0.70, with some as low as 0.38
\end{itemize}

These lower similarity values suggest that while FAISS is computationally efficient, it often compromises on retrieval quality, especially for queries requiring deeper semantic understanding.

\subsection{Game-Theoretic Outcome}

When comparing utility scores under the balanced setting of $\alpha = \beta = 1.0$, the hybrid system outperformed the FAISS-based system by a margin of 0.3679. The higher similarity scores outweighed the slightly longer query latency, highlighting that semantic accuracy contributes more significantly to overall utility than raw retrieval speed—particularly in use cases where the quality of results is paramount.

Therefore, under the given constraints and evaluation framework, the \textbf{Custom DB (Autoencoder + HNSW)} is identified as the \textbf{dominant strategy} for vector-based semantic search from a game-theoretic perspective.

\section{Conclusion}
This paper demonstrates that a game-theoretic approach to latent-space compression, leveraging deep autoencoders and hybrid HNSW indexing, significantly enhances the semantic accuracy and utility of transformer-based vector search systems compared to traditional methods like FAISS. By modeling the trade-off between retrieval accuracy and storage efficiency as a zero-sum game, our proposed framework achieves near-lossless semantic retrieval in compressed spaces, with a substantial improvement in average similarity and overall utility, albeit with a modest increase in query time. These results highlight the practical value of game-theoretic optimization for scalable, high-utility information retrieval, and pave the way for more intelligent integration of compression and search strategies in future large language model pipelines.

\bibliographystyle{splncs03_unsrt}  % LNCS uses this style
\bibliography{references_1}% references.bib file

\end{document}